\documentstyle[epsf,12pt]{article}

\begin{document}

\hoffset = -1truecm \voffset = -2truecm \baselineskip = 10 mm

\title{\bf Can a chaos term in the QCD evolution equation restrain high-energy collider physics?}

\author{
 Wei Zhu,  Zhenqi Shen and Jianhong Ruan\\\\
\normalsize Department of Physics, East China Normal University,
Shanghai 200062, P.R. China}

\date{}

\newpage

\maketitle

\vskip 3truecm

\begin{abstract}

    We indicate that the random aperiodic oscillation of the gluon
distributions in a modified BFKL equation has the positive
Lyapunov exponents. This first example of chaos in QCD evolution
equations, raises the sudden disappearance of the gluon
distributions at a critical small value of the Bjorken variable
$x$ and may stop the increase of the new particle events in a
ultra high energy hadron collider.

\end{abstract}


{\bf keywords}:  QCD evolution equation; Chaos

\newpage

    The gluon distributions are important knowledge in the
researches of high energy collider physics. The gluon
distributions of the nucleon cannot be extracted directly from the
measured structure functions in deep inelastic scattering
experiments and they mainly are predicted by using the QCD
evolution equations. However, the linear DGLAP
(Dokshitzer-Gribov-Lipatov-Altarelli-Parisi)$^{1,2}$ and BFKL
(Balitsky-Fadin-Kuraev-Lipatov)$^3$ evolution equations are no
longer reliable at ultra higher energy, therefore a series
nonlinear evolution equations, for example, the GLR-MQ-ZRS
(Gribov-Levin-Ryskin, Mueller-Qiu, Zhu-Ruan-Shen)$^{4-6}$ and BK
(Balitsky-Kovchegov)$^7$ equations were proposed, in which the
corrections of the gluon fusion are considered. An important
prediction of these equations is that the gluon distributions
approach to a so-called `saturation limit' asymptotically at small
Bjorken variable $x$, where gluon splitting is balanced with
fusion.

    As we well know, the nonlinear dynamics may have a characteristic solution--chaos,
which have been observed in many natural phenomena$^8$. Therefore,
it is interesting to ask: Does nonlinear QCD evolution equation of
gluon distributions also have chaotic solution? And how does it
impact the gluon distributions? Recently, we have proposed a
nonlinear modified BFKL (MD-BFKL) equation in ref. 9, which
describes the corrections of the gluon recombination to the BFKL
equation. We found that the unintegrated gluon distribution
function $F(x, k^2)$ in the MD-BFKL equation in the fixed coupling
approximation begins its smooth evolution under suppression of the
gluon fusion likes the solution of the BK equation, but when $x$
comes to a smaller critical $x_c$, $F(x, k^2)$ will oscillate
aperiodically near the evolution endpoint $k^2_0$. This behavior
has the characteristic feature of chaos: random and sensitivity to
the initial conditions. Furthermore, with the enhancement of
oscillation, the distribution $F(x, k^2)$ will disappear suddenly
at $x_c$. This kind of phase transition led by chaotic solutions
will certainly call our attention to reconsider the future of high
energy collider physics carefully.

    The purpose of this article is to indicate that the above
mentioned oscillate solution in the MD-BFKL equation is really
chaos and it is still holden in the running coupling case.

    We develop the MD-BFKL equation$^9$ due to the following considerations: as a
leading logarithmic approximation, the DGLAP equation neglects the
correlation of initial partons. With the increase of parton
densities, more initial gluons should be considered in evolution.
By adding initial gluons on the elementary amplitude of the DGLAP
equation in Fig. 1a step by step, we can reach the amplitudes in
Fig. 1b-1d. In a unified theoretical framework and by making use
of time ordered perturbation theory (TOPT)$^{10}$, we derived the
well-known the BFKL and GLR-MQ-ZRS equation as well as a new
MD-BFKL equation, which reads

    $$-x\frac {\partial F(x,k)}{\partial x}$$
$$=\frac{\alpha_{s}N_c}{\pi^2}\int d^2k'
\frac{k^2}{k'^2(k'-k)^2}\left[F(x,k')-\frac{1}{2}F(x,k)\right]$$
$$+\frac{18\alpha^2_s}{\pi^2R^2_N}\frac{N_c^2}{N_c^2-1}
\int d^2k' \frac{k^2}{k'^2(k'-k)^2}\left[\frac{1}{k'^2}
F^2\left(\frac{x}{2},k'\right)-\frac{1}{2k^2}F^2\left(\frac{x}{2},k\right)\right]$$
$$-\frac{36\alpha^2_s}{\pi^2R^2_N}\frac{N_c^2}{N_c^2-1}
\int d^2k' \frac{k^2}{k'^2(k'-k)^2}\left[\frac{1}{k'^2}
F^2(x,k')-\frac{1}{2k^2}F^2(x,k)\right], \eqno(1)$$ where the
three terms on the R.H.S. are the BFKL kernel of gluon splitting,
nonlinear anti-shadowing and shadowing kernels led by gluon
fusion, respectively. Each term includes real and virtual parts in
order to ensure that the evolution kernels are infrared safe. It
should be pointed out that the equation (1) is different from the
BK equation both in their elementary amplitudes and structures.
Although the derivation of this equation is tedious$^9$, the
reason of it can be proved by the consistency among this equation
and DGLAP, BFKL and GLR equations. In fact, as we see in Fig. 1,
the structures of the MD-BFKL and GLR-MQ-ZRS equations have the
similar relationship of those in the BFKL and DGLAP equations: at
leading approximation, the elementary amplitudes of the BFKL and
DGLAP equations share the same evolution kernel-gluon splitting.
But in the BFKL-initial state, there are two gluons correlated by
relative transverse momentum $k^{\prime}$, therefore the initial
state in the BFKL equation can connect with the gluon splitting
vertex by two different ways. With simple algebra, we can find
that the factor $1/k^2$ in the DGLAP evolution kernel becomes
$k^2/k^{{\prime}2}(k'-k)^2$ in the BFKL equation because of the
contributions of the interference diagrams. Similarly, the
GLR-MQ-ZRS and MD-BFKL equations have the same gluon recombination
kernel, but in the later equation the evolution kernel connect
with the initial state also by two ways. Replacing the evolution
kernel in the GLR-MQ-ZRS equation in such ways, one can reach the
MD-BFKL equation. That is, once the DGLAP, BFKL and GLR-MQ-ZRS
equations are determined, the form of the MD-BFKL equation (1) is
also fixed. We should emphasize that a complete evolution equation
must include contributions from virtual diagrams for infrared
safety, however, they are cancelled in the GLR-MQ-ZRS equation,
while cannot be neglected in the MD-BFKL equation.

    In this work, we shall use a running coupling $\alpha_s(k^2)$ in the Eq. (1)
and so we need to consider the diffusion in $\ln k^2$ with
decreasing $x$, which leads to an increasingly large contribution
from the infrared region of $k^2$ where the equation is not
expected to be valid. For this sake, as Ref. 11 we split the
integration region for real gluon emission in Eq. (1) up into two
parts: region(A) 0 to $k^2_0$ and region(B) $k^2_0$ to $\infty$.
In region(B) the MD-BFKL equation as it stands is taken to hold
and in region(A) $F(x,k^2)$ is parameterized as $C
k^2/(k^2+k^2_a)$ with $k^2_a=1 GeV^2$, where the parameter $C$
keeps the smooth connection between two parts. For simplest, we
neglect the contributions from the antishadowing effects and take
the cylindrically symmetric solution. We use the Runge-Kutta
method to compute the evolution equations. Assuming a symmetry
Gaussian input distribution exits at the evolution stating point
$x_0=10^{-3}$.

$$F(x_0=10^{-3},k^2)(k^2)^{-\frac{1}{2}}=\beta
exp\left[-\frac{\log^2 (k^2/1 GeV^2))}{40}\right], \eqno(2)$$
where $\beta=0.1$. The solutions of equation (1) are shown in
Fig.2. At first, we can see that the distribution $F(x, k^2)$
disappear suddenly at $x=x_c$. Furthermore, curves with different
values of $k^2$ turn down at same $x_c$. If we regard $x$ as an
order parameter, Fig. 2 exhibits a phase transition of first kind.

    The reason of the sudden disappearance of the gluon distribution can
be innovated by the relation of $F(x, k^2)$ and $k^2$ in Fig.3.
From the start point $x_0$, gluons diffuse on the transverse
momentum space rapidly under the action of the BFKL linear kernel.
Because of the gluon fusion, the above diffusion towards low $k^2$
is suppressed obviously. All these are like that the BK equation
has predicted. While what is different is that, when $x$ goes to
$x_c$, the oscillation of the curve will happen and increase
rapidly once it near $k^2_0$. This leads to a huge shadowing
effect and the distribution $F(x, k^2)$ disappears at $x_c$. It is
interesting that dispersed gluons will gather near $k^2_0$ before
disappearance. This kind of oscillation is random because $k$ is
not ordered in evolution. Furthermore, the above aperiodic
oscillation is very sensitive to the initial conditions.
Especially, the oscillation will be enhanced with the increase of
the numerical calculating precision. These features are also
observed in other chaos phenomena universally.

   A standard criterion of chaos is that the system has the
positive Lyapunov exponents, which indicates a strong sensitivity
to small changes in the initial conditions$^8$. We regard $y=\ln
1/x$ as `time' and calculate the Lyapunov exponents $\lambda(k^2)$
in a finite region: $10^{-7}\le x\le 0.2\times 10^{-8}$, where the
oscillation of the distribution is obvious. We divide equally the
above mentioned $y$-region into n parts with $y_1,y_2.., y_{n+1}$
and $\tau=(y_{n+1}-y_1)/n$. Assuming that the distribution evolves
to $y_1$ from $y_0=\ln 1/x_0$ and results $F(y_1,k)$.
Corresponding to a given value $F(y_1,k)$ at $(y_1,k)$, we perturb
it to $F(y_1,k)+\Delta$ with $\Delta\ll 1$. Then we continue the
evolutions from $F(y_1,k)$ and $F(y_1,k)+\Delta$ to $y_2$ from
$y_1$ respectively, and denote the resulting distributions as
$F(y_2,k)$ and $\tilde{F}(y_2,k)$. Making the difference
$\Delta_2=\vert\tilde{F}(y_2,k)-F(y_2,k)\vert$. In the following
step, we repeat the perturbation $F(y_2,k)\rightarrow
F(y_2,k)+\Delta$ and let the next evolutions from $F(y_2,k)$ and
$F(y_2,k)+\Delta$ from $y_2$ to $y_3$ respectively and get the
results $\Delta_3= \vert\tilde{F}(y_3,k)-F(y_3,k)\vert$...... (see
Fig. 4). The Lyapunov exponents for the image from $y$ to $F(y,k)$
are defined as

$$\lambda(k^2)=\lim_{n\rightarrow\infty}\frac{1}{n\tau}\sum_{i=2}^{n+1}\ln\frac {\Delta_i}
{\Delta}. \eqno(3)$$ The Lyapunov exponent of the gluon
distribution in the MD-BFKL equation with the input Eq. (2) are
presented in Fig. 5. For comparison, we give the Lyapunov
exponents but using the BFKL and BK equations in Fig. 5. The
positive values of the Lyapunov exponents clearly show that the
oscillation of $F(x, k){\sim}k^2$ is chaos of the MD-BFKL
equation. Therefore, we conclude that chaos in the MD-BFKL
equation lead to the sudden disappearance of the gluon
distributions.

    The important questions are at which scale gluons will
disappear and how much it will impact negatively on high energy
collider physics. It is a pity that we cannot predict the start
point of the evolution of MD-BFKL equation momentarily. But
considering that the MD-BFKL equation works right after GLR-MQ-ZRS
and BK equations, gluon disappearance should happen after the
saturation phenomena predicted by GLR-MQ-ZRS and BK equations.

    A typical process testing new particle with mass M on high energy hadron
collider, for example in the gluon fusion model it directly
relates to the unintegrated gluon distribution via

$$\sigma=\int d^2k_1\int d^2k_2\int_0^1dx_1\int_0^1dx_2\delta(x_1x_2-\tau)\delta^{(2)}(k_1+k_2-p_T)
F_1(x_1,k_1)F_2(x_2,k_2)d\hat{\sigma}, \eqno(4.11)$$ which is the
function of $\tau=M^2/s$. The increase of the new particle events
with increasing energy s will be stopped due to the gluon
disappearance when $\tau\le x_c$. Thus we need to re-estimate the
trials to the new physics in hadron collider physics.

    Finally, we emphasize that the MD-BFKL equation
is constructed based on a naive partonic picture (Fig. 1) and
simple leading QCD corrections. Many higher order corrections are
neglected, such as possible mixture of the operators with
different twists, the NLL (next leading logarithmic) and NLO (next
leading order) corrections, singularities from non-perturbative
parts in the factorization procedure. Of cause, the MD-BFKL
equation will only be an applicable QCD evolution equation until
all the above corrections are considered. An important questions
is: will chaos effect we demonstrated in the MD-BFKL equation
disappear after further corrections are considered? To answer this
question, we point out that chaos doesn't appear in the nonlinear
GLR equation is related to the fact that its evolution kernel has
no singularity. On the other hand, although the evolution kernel
in the nonlinear term of the BK equation is also the singular BFKL
kernel, these singularities can be absorbed into the re-definition
of the scattering amplitudes, like its form in momentum space.
Therefore the BK equation has no chaotic solution, either. So we
suggest the fact that chaos appear in the MD-BFKL equation firstly
is related to the singularities in its nonlinear terms. From the
experiences in the study of the BFKL kernel, those possible QCD
corrections we mentioned to the MD-BFKL will probably make
singularities of this nonlinear equation more complicated, instead
of removing these singularities completely, since the regularized
singular parts of the evolution kernel always dominate the
evolution. In this situation, we could expect that more
interesting chaos phenomena will appear in the new MD-BFKL
equation. These phenomena will be most interested to high energy
physics and nonlinear science.

    In summary, a random oscillation of the unintegrated gluon
distributions in a modified BFKL equation is indicated as chaos,
which has the positive Lyapunov exponents. This first example of
chaos in QCD evolution equations, raises the sudden disappearance
of the gluon distributions at a critical small value of the
Bjorken variable $x$ and may stop the increase of the new particle
events in a ultra high energy hadron collider.

\vspace{0.3cm}

\noindent {\bf Acknowledgments}:  We thank Z.H. Liu for useful
discussions in chaos. This work was supported by National Natural
Science Foundations of China 10475028.

\newpage

Figure Captions

Fig. 1 A part of the amplitudes for four related evolution
equations. (a) The DGLAP equation; (b) the BFKL equation; (c and
d) the GLR-MQ-ZRS and MD-BFKL equations, which contain the leading
corrections of the gluon recombination to the DGLAP and BFKL
equations, respectively.

Fig.2.  x-dependence of the unintegrated gluon distribution in the
MD-BFKL equation (1) for different values of $k^2$. The results
show that $F(x,k^2)$ suddenly drops near a critical value of
$x\sim x_c\simeq 3.8\times 10^{-8}$. The dashed curves are the
corresponding solution of the BFKL equation with $k^2=50GeV^2$.

Fig.3. $k^2$-dependence of the unintegrated gluon distribution for
different values of $x$. The results present the oscillations near
$x_c$ when $k^2\rightarrow k^2_0$. This leads to a huge shadowing
effect and the distribution $F(x, k^2)$ disappears at $x<x_c$.

Fig. 4. Schematic programs calculating the
Lyapunov~exponents~of~the~evolution~equation. The results are
insensitive to the value of $\Delta$.

Fig. 5 Plots of the Lyapunov exponents in the region $10^{-7}\le
x\le 0.2\times 10^{-8}$. The positive Lyapunov exponents show that
the corresponding solution of the MD-BFKL equation is chaos.

\newpage
\epsfysize=22cm\epsffile{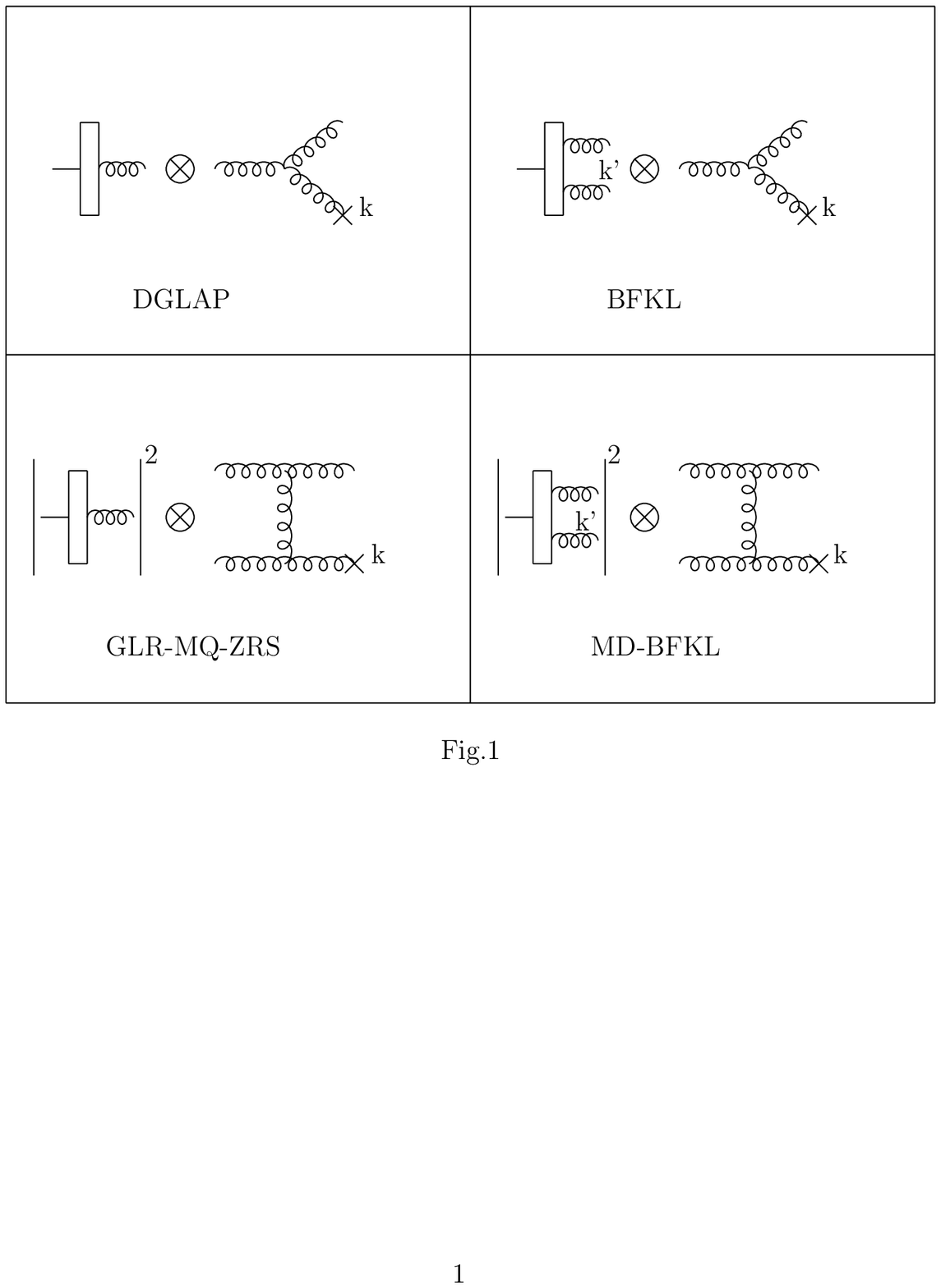}
\newpage
\epsfysize=22cm\epsffile{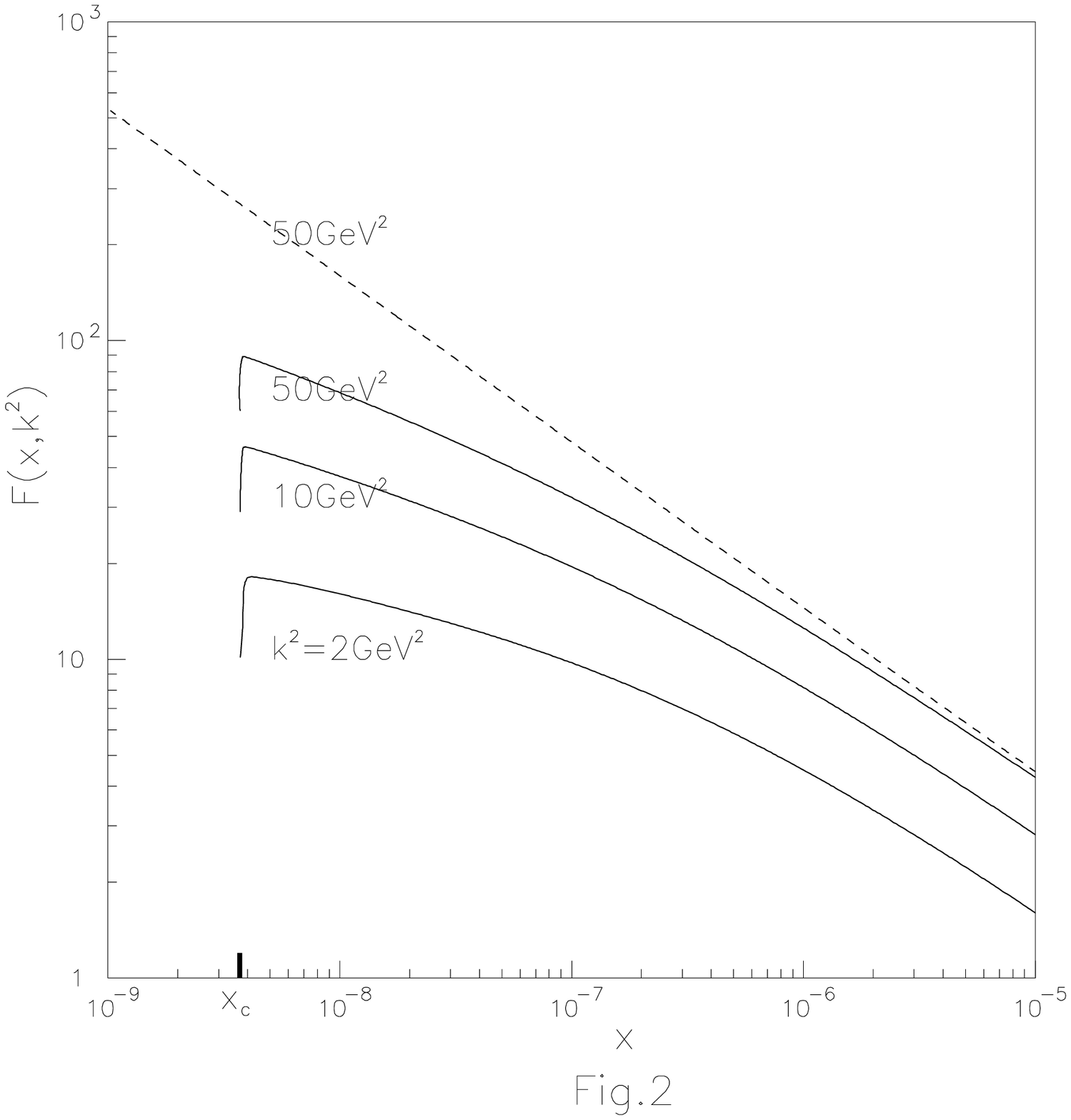}
\newpage
\epsfysize=22cm\epsffile{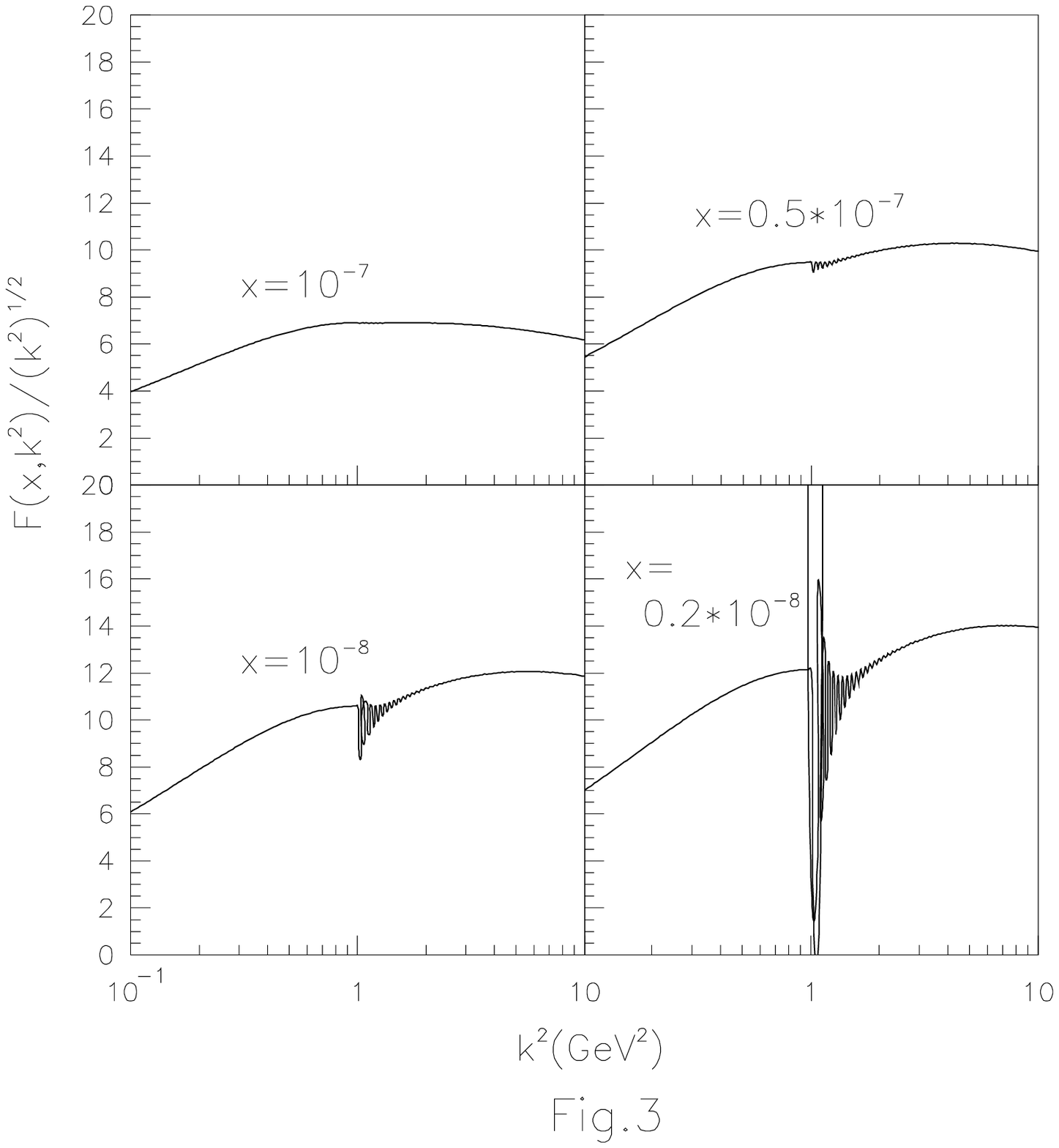}
\newpage
\epsfysize=22cm\epsffile{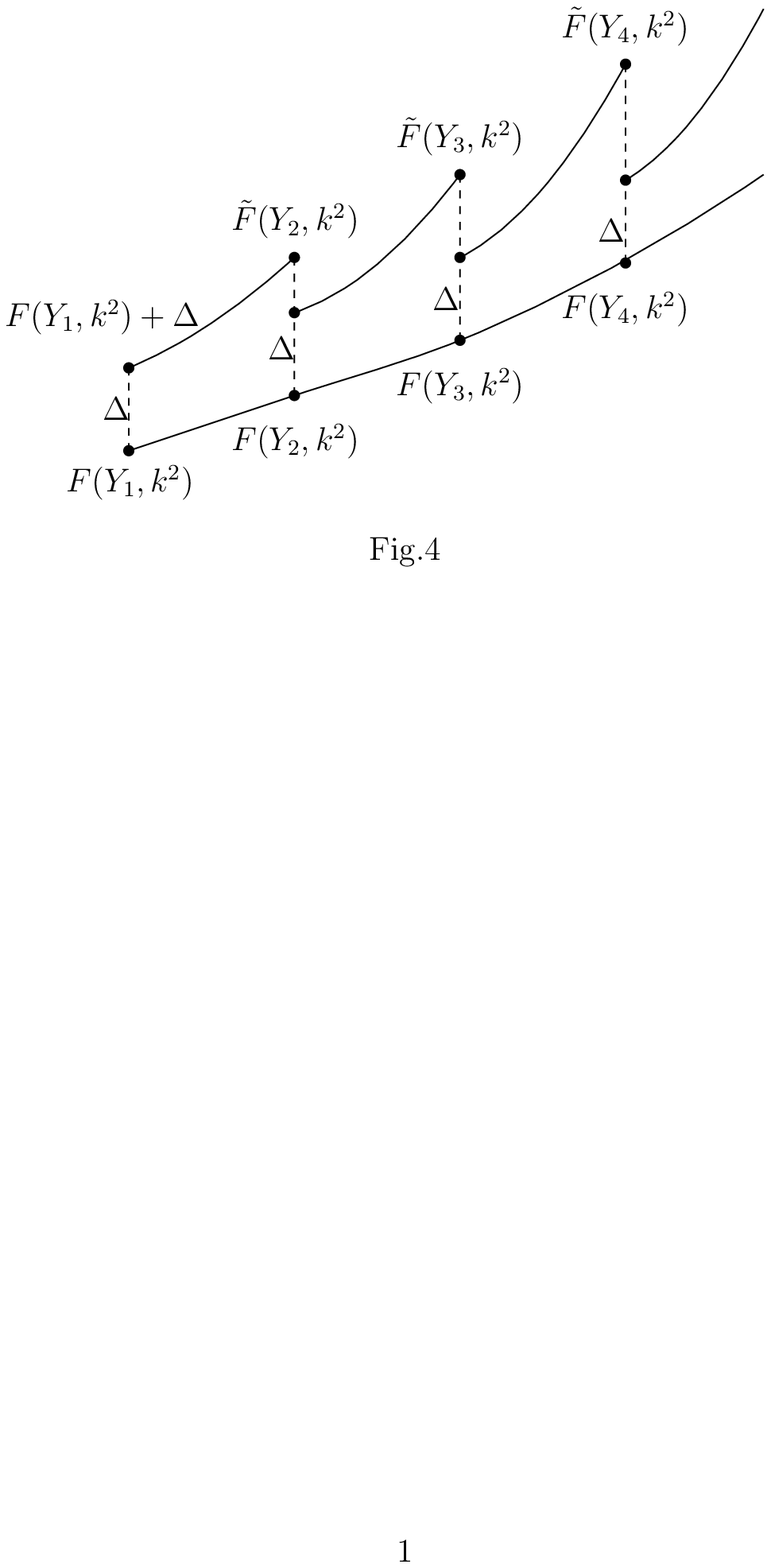}
\newpage
\epsfysize=22cm\epsffile{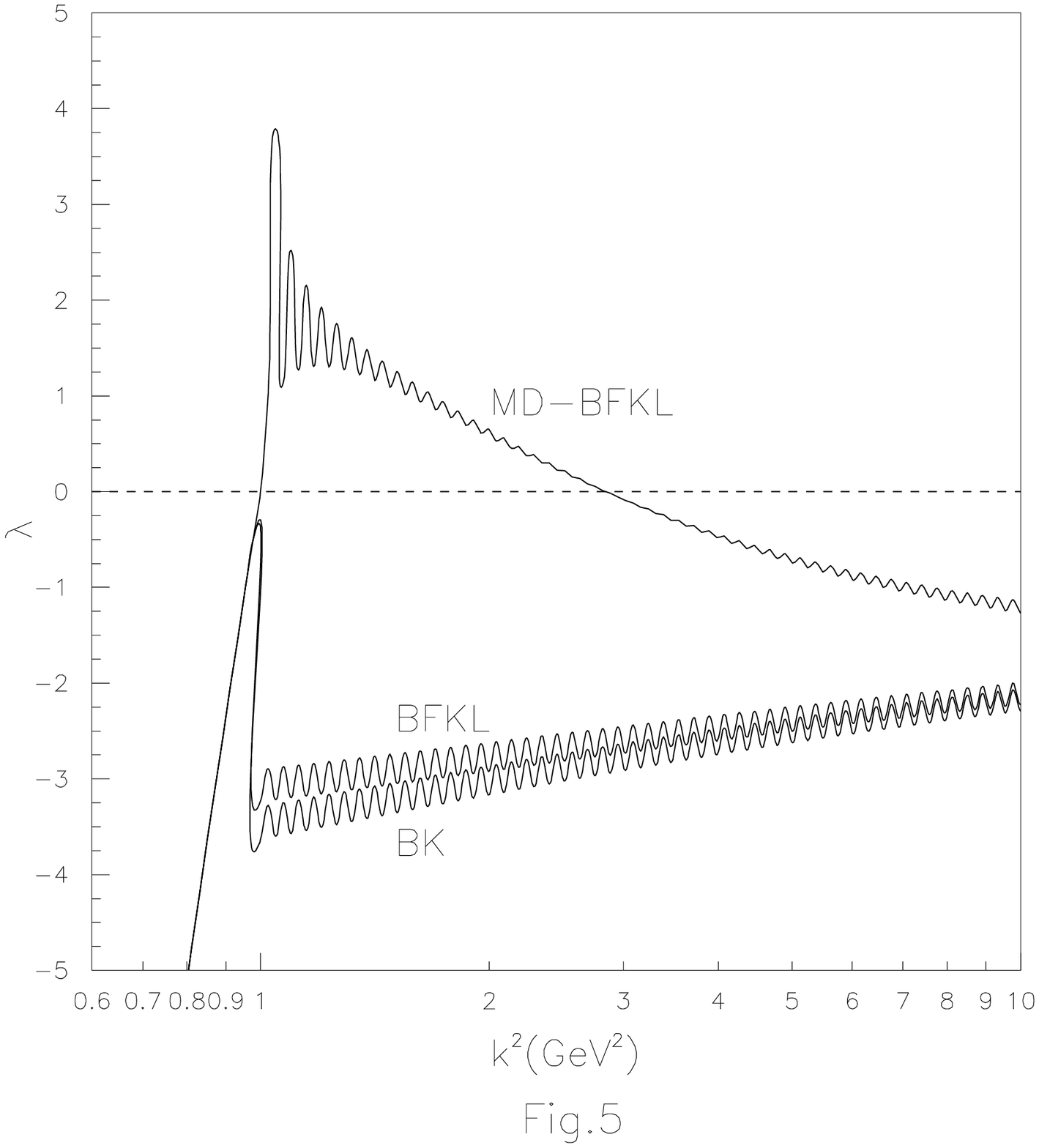}

\end{document}